\documentstyle[12pt]{article}

\catcode`\@=11
\long\def\@makefntext#1{ 
\protect\noindent \hbox to 3.2pt {\hskip-.9pt
$^{{\ninerm\@thefnmark}}$\hfil}#1\hfill} 

\def\thefootnote{\fnsymbol{footnote}}
 \def\@makefnmark{\hbox to 0pt{$^{\@thefnmark}$\hss}}  

\def\ps@myheadings{\let\@mkboth\@gobbletwo
\def\@oddhead{\hbox{} 
\rightmark\hfil\ninerm\thepage}
\def\@oddfoot{}\def\@evenhead{\ninerm\thepage\hfil 
\leftmark\hbox{}}\def\@evenfoot{}
\def\sectionmark##1{}\def\subsectionmark##1{}}

\def\CVector#1#2#3{
\left\vert\begin{array}{c}
                 #1 \\
                 #2
                \end{array}, #3\right>}

\def\AVector#1#2#3{\left|  #1, #2; #3 \right>}

\begin{document}

\newcommand{\symbolfootnote}{\renewcommand{\thefootnote}
        {\fnsymbol{footnote}}}
\renewcommand{\thefootnote}{\fnsymbol{footnote}}
\newcommand{\alphfootnote}
        {\setcounter{footnote}{0}
         \renewcommand{\thefootnote}{\sevenrm\alph{footnote}}}

\newcounter{sectionc}\newcounter{subsectionc}\newcounter{subsubsectionc}
\renewcommand{\section}[1] {\vspace{0.6cm}\addtocounter{sectionc}{1}
\setcounter{subsectionc}{0}\setcounter{subsubsectionc}{0}\noindent
        {\bf\thesectionc. #1}\par\vspace{0.4cm}}
\renewcommand{\subsection}[1] {\vspace{0.6cm}\addtocounter{subsectionc}{1}
        \setcounter{subsubsectionc}{0}\noindent
        {\it\thesectionc.\thesubsectionc. #1}\par\vspace{0.4cm}}
\renewcommand{\subsubsection}[1]
{\vspace{0.6cm}\addtocounter{subsubsectionc}{1}
        \noindent {\rm\thesectionc.\thesubsectionc.\thesubsubsectionc.
        #1}\par\vspace{0.4cm}}
\newcommand{\nonumsection}[1] {\vspace{0.6cm}\noindent{\bf #1}
        \par\vspace{0.4cm}}

\newcounter{appendixc}
\newcounter{subappendixc}[appendixc]
\newcounter{subsubappendixc}[subappendixc]
\renewcommand{\thesubappendixc}{\Alph{appendixc}.\arabic{subappendixc}}
\renewcommand{\thesubsubappendixc}
        {\Alph{appendixc}.\arabic{subappendixc}.\arabic{subsubappendixc}}

\renewcommand{\appendix}[1] {\vspace{0.6cm}
        \refstepcounter{appendixc}
        \setcounter{figure}{0}
        \setcounter{table}{0}
        \setcounter{equation}{0}
        \renewcommand{\thefigure}{\Alph{appendixc}.\arabic{figure}}
        \renewcommand{\thetable}{\Alph{appendixc}.\arabic{table}}
        \renewcommand{\theappendixc}{\Alph{appendixc}}
        \renewcommand{\theequation}{\Alph{appendixc}.\arabic{equation}}
        \noindent{\bf Appendix \theappendixc #1}\par\vspace{0.4cm}}
\newcommand{\subappendix}[1] {\vspace{0.6cm}
        \refstepcounter{subappendixc}
        \noindent{\bf Appendix \thesubappendixc. #1}\par\vspace{0.4cm}}
\newcommand{\subsubappendix}[1] {\vspace{0.6cm}
        \refstepcounter{subsubappendixc}
        \noindent{\it Appendix \thesubsubappendixc. #1}
        \par\vspace{0.4cm}}

\def\abstracts#1{{
        \centering{\begin{minipage}{30pc}\tenrm\baselineskip=12pt\noindent
        \centerline{\tenrm ABSTRACT}\vspace{0.3cm}
        \parindent=0pt #1
        \end{minipage} }\par}}

\newcommand{\bibit}{\it}
\newcommand{\bibbf}{\bf}
\renewenvironment{thebibliography}[1]
        {\begin{list}{\arabic{enumi}.}
       {\usecounter{enumi}\setlength{\parsep}{0pt}
\setlength{\leftmargin 1.25cm}{\rightmargin 0pt}
         \setlength{\itemsep}{0pt} \settowidth
        {\labelwidth}{#1.}\sloppy}}{\end{list}}

\topsep=0in\parsep=0in\itemsep=0in
\parindent=1.5pc

\newcounter{itemlistc}
\newcounter{romanlistc}
\newcounter{alphlistc}
\newcounter{arabiclistc}
\newenvironment{itemlist}
        {\setcounter{itemlistc}{0}
         \begin{list}{$\bullet$}
        {\usecounter{itemlistc}
         \setlength{\parsep}{0pt}
         \setlength{\itemsep}{0pt}}}{\end{list}}

\newenvironment{romanlist}
        {\setcounter{romanlistc}{0}
         \begin{list}{$($\roman{romanlistc}$)$}
        {\usecounter{romanlistc}
         \setlength{\parsep}{0pt}
         \setlength{\itemsep}{0pt}}}{\end{list}}

\newenvironment{alphlist}
        {\setcounter{alphlistc}{0}
         \begin{list}{$($\alph{alphlistc}$)$}
        {\usecounter{alphlistc}
         \setlength{\parsep}{0pt}
         \setlength{\itemsep}{0pt}}}{\end{list}}

\newenvironment{arabiclist}
        {\setcounter{arabiclistc}{0}
         \begin{list}{\arabic{arabiclistc}}
        {\usecounter{arabiclistc}
         \setlength{\parsep}{0pt}
         \setlength{\itemsep}{0pt}}}{\end{list}}

\newcommand{\fcaption}[1]{
        \refstepcounter{figure}
        \setbox\@tempboxa = \hbox{\tenrm Fig.~\thefigure. #1}
        \ifdim \wd\@tempboxa > 6in
           {\begin{center}
        \parbox{6in}{\tenrm\baselineskip=12pt Fig.~\thefigure. #1 }
            \end{center}}
        \else
             {\begin{center}
             {\tenrm Fig.~\thefigure. #1}
              \end{center}}
        \fi}

\newcommand{\tcaption}[1]{
        \refstepcounter{table}
        \setbox\@tempboxa = \hbox{\tenrm Table~\thetable. #1}
        \ifdim \wd\@tempboxa > 6in
           {\begin{center}
        \parbox{6in}{\tenrm\baselineskip=12pt Table~\thetable. #1 }
            \end{center}}
        \else
             {\begin{center}
             {\tenrm Table~\thetable. #1}
              \end{center}}
        \fi}

\def\@citex[#1]#2{\if@filesw\immediate\write\@auxout
        {\string\citation{#2}}\fi
\def\@citea{}\@cite{\@for\@citeb:=#2\do
        {\@citea\def\@citea{,}\@ifundefined
        {b@\@citeb}{{\bf ?}\@warning
        {Citation `\@citeb' on page \thepage \space undefined}}
        {\csname b@\@citeb\endcsname}}}{#1}}

\newif\if@cghi
\def\cite{\@cghitrue\@ifnextchar [{\@tempswatrue
        \@citex}{\@tempswafalse\@citex[]}}
\def\citelow{\@cghifalse\@ifnextchar [{\@tempswatrue
        \@citex}{\@tempswafalse\@citex[]}}
\def\@cite#1#2{{$\null^{#1}$\if@tempswa\typeout
        {IJCGA warning: optional citation argument
        ignored: `#2'} \fi}}
\newcommand{\citeup}{\cite}

\def\fnm#1{$^{\mbox{\scriptsize #1}}$}
\def\fnt#1#2{\footnotetext{\kern-.3em
        {$^{\mbox{\sevenrm #1}}$}{#2}}}

\font\twelvebf=cmbx10 scaled\magstep 1
\font\twelverm=cmr10 scaled\magstep 1
\font\twelveit=cmti10 scaled\magstep 1
\font\elevenbfit=cmbxti10 scaled\magstephalf
\font\elevenbf=cmbx10 scaled\magstephalf
\font\elevenrm=cmr10 scaled\magstephalf
\font\elevenit=cmti10 scaled\magstephalf
\font\bfit=cmbxti10
\font\tenbf=cmbx10
\font\tenrm=cmr10
\font\tenit=cmti10
\font\ninebf=cmbx9
\font\ninerm=cmr9
\font\nineit=cmti9
\font\eightbf=cmbx8
\font\eightrm=cmr8
\font\eightit=cmti8


\centerline{\tenbf QUANTUM ALGEBRAIC SYMMETRIES IN NUCLEI AND MOLECULES}
\vspace{0.8cm}
\centerline{\tenrm Dennis Bonatsos$^{1,2}$, C. Daskaloyannis$^3$,
P. Kolokotronis$^2$ and D. Lenis$^2$}
\baselineskip=13pt
\centerline{\tenit $^1$ ECT$^*$, Villa Tambosi, Strada delle Tabarelle 286}
\baselineskip=12pt
\centerline{\tenit I-38050 Villazzano (Trento), Italy}
\baselineskip=13pt
\centerline{\tenit $^2$ Institute of Nuclear Physics, NCSR ``Demokritos''}
\baselineskip=12pt
\centerline{\tenit GR-15310 Aghia Paraskevi, Attiki, Greece}
\baselineskip=13pt
\centerline{\tenit $^3$ Department of Physics, Aristotle University of
Thessaloniki}
\baselineskip=12pt
\centerline{\tenit GR-54006 Thessaloniki, Greece}
\vspace{0.9cm}
\abstracts{Various applications of quantum algebraic techniques in nuclear
and molecular physics
 are briefly reviewed. Emphasis is put in the study of
the symmetries of the anisotropic quantum harmonic oscillator with rational
ratios of frequencies, which underly the structure of superdeformed and
hyperdeformed nuclei, the Bloch--Brink $\alpha$-cluster model and possibly
the shell structure in deformed atomic clusters.
}

\vfil
\rm\baselineskip=12pt
\section{Introduction}

Quantum algebras [1,2] (also called quantum groups) are deformed versions
of the
usual Lie algebras, to which they reduce when the deformation parameter
$q$ is set equal to unity. Their use in physics became popular with the
introduction [3--5] of the $q$-deformed harmonic oscillator as a tool for
providing a boson realization of the quantum algebra su$_q$(2), although
similar mathematical structures had already been known [6,7].
Initially used for solving the quantum Yang--Baxter equation, quantum algebras
have subsequently found applications in several branches of physics, as, for
example, in the description of spin chains, squeezed states, rotational
and vibrational nuclear and molecular spectra, and in conformal
field theories. By now several kinds of generalized deformed oscillators [8,9]
and generalized deformed su(2) algebras [10--12]  have been introduced.

The main applications of quantum algebraic techniques in nuclear physics
include:

i) The su$_q$(2) rotator model, used for the description of rotational
spectra of deformed [13,14] and superdeformed [15] nuclei,
as well as for the description
of the electromagnetic transition probabilities [16] connecting the energy
levels within each band.

ii) Extensions of the su$_q$(2) model suitable for the description of
vibrational and transitional nuclear spectra [17].

iii)  Generalized deformed su$_{\Phi}$(2) algebras [10], characterized by a
structure function $\Phi$ and possessing a representation theory similar
to that of the usual su(2) algebra. The su$_q$(2) algebra occurs for a
special choice of $\Phi$, while different choices lead to different
energy formulae for the rotational spectrum, as, for example, the
Holmberg--Lipas formula [18].

iv) The use of deformed bosons for the description of pairing correlations
in nuclei [19--21].

v) The construction of deformed versions of various exactly soluble models,
as the Interacting Boson Model [22] and the Moszkowski model [23]. In the
framework of the latter, RPA modes [24] and high spin states [25] have also
been studied.

vi) The study of the symmetries of the anisotropic quantum harmonic
oscillator with rational ratios of frequencies (RHO), which are of interest
in connection with superdeformed and hyperdeformed nuclei [26,27], ``pancake''
nuclei (i.e. very oblate nuclei) [27], cluster configurations in light nuclei
[28] connected to the Bloch--Brink $\alpha$ cluster model [29],
and possibly in
connection with deformed atomic clusters [30,31].

The main applications of quantum algebraic techniques in molecular physics
include:

i) The su$_q$(2) rotator model, used for the description of rotational
spectra of diatomic molecules [32].

ii) The description of vibrational spectra of diatomic molecules in terms
of the su$_q$(1,1) algebra [33,34], as well as in terms of generalized deformed
oscillators [35]. WKB equivalent potentials possessing the same spectrum
as these oscillators have also been constucted [36]. These oscillators have
been found appropriate for approximating certain Quasi Exactly Soluble
Potentials [37].

iii) The development of a deformed version of the vibron model [38] of
molecular structure.

iv) The construction of deformed oscillators equivalent to the Morse
potential [39] and the use of a set of such oscillators for the description
of the vibrational spectra of highly symmetric polyatomic molecules [40].

In the remainder of this paper
 the symmetries of the anisotropic quantum harmonic oscillator
with rational ratios of frequencies will be considered in more detail.

\section{ Anisotropic quantum harmonic oscillator with rational ratios of
frequencies}

The symmetries of the 3-dimensional anisotropic quantum harmonic oscillator
with rational ratios of frequencies (RHO) are of high current interest in
nuclear physics, since they are
the basic symmetries  underlying the structure of superdeformed and
hyperdeformed nuclei [26,27].
The 2-dimensional RHO is also of interest, in
connection with ``pancake'' nuclei, i.e. very oblate nuclei [27]. Cluster
configurations in light nuclei can also be  described in terms of RHO
symmetries [28], which underlie the geometrical structure of the
Bloch--Brink $\alpha$-cluster model [29]. The 3-dim RHO is also of
interest for the interpretation
of the observed shell structure in atomic clusters [30], especially
after the
realization [31] that large deformations can occur in such systems.

The two-dimensional [41,42] and
three-dimensional [43,44]  anisotropic harmonic
oscillators have been the subject of several investigations, both at the
classical and the quantum mechanical level. These oscillators are examples
of superintegrable systems [45]. The special cases with frequency ratios
1:2 [46] and 1:3 [47] have also been considered.
While
at the classical level it is clear that the su(N) or sp(2N,R) algebras can
be used for the description of the N-dimensional anisotropic oscillator, the
situation at the quantum level, even in the two-dimensional case, is not as
simple.

In the remainder of this section the 2-dim RHO will be considered in more
detail.

\subsection{ The deformed u(2) algebra}

Let us consider the system described by the Hamiltonian:
\begin{equation}
\label{eq:Hamiltonian}H=\frac{1}{2}\left( {p_x}^2 + {p_y}^2 + \frac{x^2}{m^2}
+ \frac{y^2}{n^2} \right),
\end{equation}
where $m$ and $n$ are two natural numbers mutually prime ones, i.e. their
great common divisor is $\gcd (m,n)=1$.

We define the creation and annihilation operators [41]
\begin{equation}
\label{eq:operators}
a^\dagger=\frac{x/m - i p_x}{\sqrt{2}}, \quad a =
\frac{x/m + i p_x}{\sqrt{2}}, \qquad b^\dagger=\frac{y/n - i p_y}{\sqrt{%
2}}, \quad b=\frac{y/n + i p_y}{\sqrt{2}}.
\end{equation}
These operators satisfy the commutation relations:
\begin{equation}
\label{eq:commutators}\left[ a,a^\dagger \right] = \frac{1}{m}, \quad \left[
b,b^\dagger \right] = \frac{1}{n}, \quad \mbox{other commutators}=0.
\end{equation}
One can further define
$$
U=\frac{1}{2} \left\{ a, a^\dagger \right\}, \qquad W=\frac{1}{2} \left\{ b,
b^\dagger \right\}.%
$$
One  can then define the enveloping algebra generated by the operators:
\begin{equation}
\label{eq:generators}
\begin{array}{c}
S_+= \left(a^\dagger\right)^m \left(b\right)^n,\quad S_-= \left(a\right)^m
\left(b^\dagger\right)^n, \\
[0.24in] S_0= \frac{1}{2}\left( U - W \right), \quad H=U+W.
\end{array}
\end{equation}
These genarators satisfy the following relations:
\begin{equation}
\label{eq:SS}\left[ S_0,S_\pm \right]=\pm S_\pm, \quad \left[H,S_i\right]=0,
\quad \mbox{for}\quad i=0,\pm,
\end{equation}
and
$$
S_+S_- = \prod\limits_{k=1}^{m}\left( U - \frac{2k-1}{2m} \right)
\prod\limits_{\ell=1}^{n}\left( W + \frac{2\ell-1}{2n} \right),
$$
$$
S_-S_+ = \prod\limits_{k=1}^{m}\left( U + \frac{2k-1}{2m} \right)
\prod\limits_{\ell=1}^{n}\left( W - \frac{2\ell-1}{2n} \right).
$$
The fact that the operators $S_i$, $i=0, \pm$ are integrals of motion has
been already realized in [41].

The above relations mean that the harmonic oscillator of Eq. (\ref
{eq:Hamiltonian}) is described by the enveloping algebra of the
generalization of the u(2) algebra formed by the generators $S_0$, $S_+$, $%
S_-$ and $H$, satisfying the commutation relations of Eq. (\ref{eq:SS}) and
\begin{equation}
\label{eq:U2}
\begin{array}{c}
\left[S_-,S_+\right] = F_{m,n} (H,S_0+1) - F_{m,n} (H,S_0), \\
[0.24 in] \mbox{where}\quad F_{m,n}(H,S_0)= \prod\limits_{k=1}^{m}\left(
H/2+S_0 - \frac{2k-1}{2m} \right) \prod\limits_{\ell=1}^{n}\left( H/2-S_0 +
\frac{2\ell-1}{2n} \right).
\end{array}
\end{equation}
In the case of $m=1$, $n=1$ this algebra is the usual u(2) algebra, and the
operators $S_0,S_\pm$ satisfy the commutation relations of the ordinary u(2)
algebra, since in this case one easily finds that
$$
[S_-, S_+]=-2 S_0.%
$$
In the rest of the cases, the algebra is a deformed version of u(2), in
which the commutator $[S_-,S_+]$ is a polynomial of $S_0$ of order $m+n-1$.

\subsection{ The representations}

The finite dimensional representation modules  of this algebra can be found
using the concept of the generalized deformed oscillator [8], in a
method similar to the one used in [48]  for the study of quantum
superintegrable systems. The operators:
\begin{equation}
\label{eq:alge-gen}{\cal A}^\dagger= S_+, \quad {\cal A}= S_-, \quad {\cal N}%
= S_0-u, \quad u=\mbox{ constant},
\end{equation}
where $u$ is a constant to be determined, are the generators of a deformed
oscillator algebra:
$$
\left[ {\cal N} , {\cal A}^\dagger \right] = {\cal A}^\dagger, \quad \left[
{\cal N} , {\cal A} \right] = -{\cal A}, \quad {\cal A}^\dagger{\cal A}
=\Phi( H, {\cal N} ), \quad {\cal A}{\cal A}^\dagger =\Phi( H, {\cal N}+1 ).
$$
The structure function $\Phi$ of this algebra is determined by the function $%
F_{m,n}$ in Eq. (\ref{eq:U2}):
\begin{equation}
\label{eq:sf}
\begin{array}{l}
\Phi( H,
{\cal N} )= F_{m,n} (H,{\cal N} +u ) = \\ = \prod\limits_{k=1}^{m}\left( H/2+%
{\cal N} +u - \frac{2k-1}{2m} \right) \prod\limits_{\ell=1}^{n}\left( H/2-%
{\cal N} - u + \frac{2\ell-1}{2n} \right).
\end{array}
\end{equation}
The deformed oscillator corresponding to the structure function of Eq.  (\ref
{eq:sf}) has an energy dependent Fock space of dimension $N+1$ if
\begin{equation}
\label{eq:equations}\Phi(E,0)=0, \quad \Phi(E, N+1)=0, \quad \Phi(E,k)>0,
\quad \mbox{for} \quad k=1,2,\ldots,N.
\end{equation}
The Fock space is defined by:
\begin{equation}
H\vert E, k > =E \vert E, k >, \quad {\cal N} \vert E, k >= k \vert E, k
>,\quad a\vert E, 0 >=0,
\end{equation}
\begin{equation}
{\cal A}^\dagger \vert E, k> = \sqrt{\Phi(E,k+1)} \vert E, k+1>, \quad {\cal %
A} \vert E, k> = \sqrt{\Phi(E,k)} \vert E, k-1>.
\end{equation}
The basis of the Fock space is given by:
$$
\vert E, k >= \frac{1}{\sqrt{[k]!}} \left({\cal A}^\dagger\right)^k\vert E,
0 >, \quad k=0,1,\ldots N,
$$
where the ``factorial'' $[k]!$ is defined by the recurrence relation:
$$
[0]!=1, \quad [k]!=\Phi(E,k)[k-1]! \quad .
$$
Using the Fock basis we can find the matrix representation of the deformed
oscillator and then the matrix representation of the algebra of Eqs (\ref
{eq:SS}), (\ref{eq:U2}). The solution of Eqs (\ref{eq:equations}) implies
the following pairs of permitted values for the energy eigenvalue $E$ and
the constant $u$:
\begin{equation}
\label{eq:E1}E=N+\frac{2p-1}{2m}+\frac{2q-1}{2n} ,
\end{equation}
where $p=1,2,\ldots,m$, $q=1,2,\ldots,n$, and
$$
u=\frac{1}{2}\left( \frac{2p-1}{2m}-\frac{2q-1}{2n} -N \right),
$$
the corresponding structure function being given by:
\begin{equation}
\label{eq:structure-function}
\begin{array}{l}
\Phi(E,x)=\Phi^{N}_{(p,q)}(x)= \\
=\prod\limits_{k=1}^{m}\left( x +
\displaystyle \frac{2p-1}{2m}- \frac{2k-1}{2m} \right)
\prod\limits_{\ell=1}^{n}\left( N-x+ \displaystyle \frac{2q-1}{2n} + \frac{%
2\ell-1}{2n}\right) \\ =\displaystyle\frac{1}{m^m n^n} \displaystyle\frac{
\Gamma\left(mx+p\right) }{\Gamma\left(mx+p-m\right)} \displaystyle \frac{%
\Gamma\left( (N-x)n + q + n \right)} {\Gamma\left( (N-x)n + q \right)}.
\end{array}
\end{equation}
In all these equations one has $N=0,1,2,\ldots$, while the dimensionality of
the representation is given by $N+1$. Eq. (\ref{eq:E1})  means that there
are $m\cdot n$ energy eigenvalues corresponding to each $N$ value, each
eigenvalue having degeneracy $N+1$. (Later we shall see that the degenerate
states corresponding to the same eigenvalue can be labelled by an ``angular
momentum''.)

The energy formula can be corroborated by using the
corresponding Schr\"{o}dinger equation. For the Hamiltonian of Eq. (\ref
{eq:Hamiltonian}) the eigenvalues of the Schr\"{o}dinger equation are given
by:
\begin{equation}
\label{eq:E2}E=\frac{1}{m}\left(n_x+\frac{1}{2}\right)+ \frac{1}{n}\left(n_y+%
\frac{1}{2}\right),
\end{equation}
where $n_x=0,1,\ldots$ and $n_y=0,1,\ldots$. Comparing Eqs (\ref{eq:E1}) and
(\ref{eq:E2}) one concludes that:
$$
N= \left[n_x/m\right]+\left[n_y/n\right],%
$$
where $[x]$ is the integer part of the number $x$, and
$$
p=\mbox{mod}(n_x,m)+1, \quad q=\mbox{mod}(n_y,n)+1.
$$

The eigenvectors of the Hamiltonian can be parametrized by the
dimensionality of the representation $N$, the numbers $p,q$, and the number $%
k=0,1,\ldots,N$. $k$ can be identified as $[n_x/m]$. One then has:

\begin{equation}
\label{eq:en-rep}H\left\vert
\begin{array}{c}
N \\
(p,q)
\end{array}
, k \right>= \left(N+\displaystyle
\frac{2p-1}{2m}+\frac{2q-1}{2n} \right)\left\vert
\begin{array}{c}
N \\
(p,q)
\end{array}
, k \right>,
\end{equation}
\begin{equation}
\label{eq:s0-rep}S_0 \left\vert
\begin{array}{c}
N \\
(p,q)
\end{array}
, k \right>= \left( k+ \displaystyle
\frac{1}{2} \left( \frac{2p-1}{2m}- \frac{2q-1}{2n} -N \right) \right)
\left\vert
\begin{array}{c}
N \\
(p,q)
\end{array}
, k \right>,
\end{equation}
\begin{equation}
\label{eq:sp-rep}S_+\left\vert
\begin{array}{c}
N \\
(p,q)
\end{array}
, k \right> = \sqrt{ \Phi^N_{(p,q)}(k+1)} \left\vert
\begin{array}{c}
N \\
(p,q)
\end{array}
, k +1\right>,
\end{equation}
\begin{equation}
\label{eq:sm-rep}S_-\left\vert
\begin{array}{c}
N \\
(p,q)
\end{array}
, k \right> = \sqrt{ \Phi^N_{(p,q)}(k)} \left\vert
\begin{array}{c}
N \\
(p,q)
\end{array}
, k -1\right>.
\end{equation}

\subsection{ The ``angular momentum'' quantum number}

It is worth noticing that the operators $S_0,S_\pm$ do not correspond to a
generalization of the angular momentum, $S_0$ being the operator
corresponding to the Fradkin operator $S_{xx}-S_{yy}$ [49].
The corresponding ``angular momentum'' is defined by:
\begin{equation}
\label{eq:angular-momentum}L_0=-i\left(S_+-S_-\right).
\end{equation}
The ``angular momentum'' operator commutes with the Hamiltonian:
$$
\left[ H,L_0 \right]=0.
$$
Let $\vert \ell> $ be the eigenvector of the operator $L_0$ corresponding to
the eigenvalue $\ell$. The general form of this eigenvector can be given by:
\begin{equation}
\vert \ell > = \sum\limits_{k=0}^N \frac{i^k c_k}{\sqrt{[k]!}} \left\vert
\begin{array}{c}
N \\
(p,q)
\end{array}
, k \right>.
\label{eq:ang-vector}
\end{equation}

In order to find the eigenvalues of $L$ and the coefficients $c_k$ we use
the Lanczos algorithm, as formulated in [50]. From Eqs (\ref
{eq:sp-rep}) and (\ref{eq:sm-rep}) we find
$$
\begin{array}{l}
L_0|\ell >=\ell |\ell >=\ell \sum\limits_{k=0}^N\frac{i^kc_k}{\sqrt{[k]!}}%
\left|
\begin{array}{c}
N \\
(p,q)
\end{array}
,k\right\rangle = \\
=\frac 1i\sum\limits_{k=0}^{N-1}\frac{i^kc_k\sqrt{\Phi _{(p,q)}^N(k+1)}}{%
\sqrt{[k]!}}\left|
\begin{array}{c}
N \\
(p,q)
\end{array}
,k+1\right\rangle -\frac 1i\sum\limits_{k=1}^N\frac{i^kc_k\sqrt{\Phi
_{(p,q)}^N(k)}}{\sqrt{[k]!}}\left|
\begin{array}{c}
N \\
(p,q)
\end{array}
,k-1\right\rangle
\end{array}
$$
{}From this equation we find that:
$$
c_k=(-1)^k 2^{-k/2}H_k(\ell /\sqrt{2})/{\cal N},
\quad {\cal N}^2= \sum\limits_{n=0}^N 2^{-n}H_n^2(\ell /\sqrt{2})
$$
where the function $H_k(x)$ is a generalization of the ``Hermite''
polynomials (see also [51,52]), satisfying the recurrence
relations:
$$
H_{-1}(x)=0,\quad H_0(x)=1,
$$
$$
H_{k+1}(x)=2xH_k(x)-2\Phi _{(p,q)}^N(k)H_{k-1}(x),
$$
and the ``angular momentum'' eigenvalues $\ell $ are the roots of the
polynomial equation:
\begin{equation}
H_{N+1}(\ell /\sqrt{2})=0.
\label{eq:eigenvalues}
\end{equation}
Therefore for a given value of $N$ there are $N+1$ ``angular momentum''
eigenvalues $\ell $, symmetric around zero (i.e. if $\ell $ is an ``angular
momentum'' eigenvalue, then $-\ell $ is also an ``angular momentum''
eigenvalue). In the case of the symmetric harmonic oscillator ($m/n=1/1$)
these eigenvalues are uniformly distributed and differ by 2. In the general
case the ``angular momentum'' eigenvalues are non-uniformly distributed. For
small values of $N$ analytical formulae for the ``angular momentum''
eigenvalues can be found [51]. Remember that to each value of $N$
correspond $m\cdot n$ energy levels, each with degeneracy $N+1$.

In order to have a formalism corresponding to the one of the isotropic
oscillator, let us introduce  for every $N$ and
$(p,q)$ an ordering of the ``angular momentum'' eigenvalues
$$
\ell_m^{L,(p,q)}, \quad \mbox{where} \quad L=N
\quad \mbox{and} \quad m=-L,-L+2,\ldots,L-2,L,
$$
by assuming that:
$$
\ell_m^{L,(p,q)} \le \ell_n^{L,(p,q)} \quad \mbox{if} \quad m<n,
$$
 the corresponding eigenstate being given by:
\begin{equation}
\AVector{L}{m}{(p,q)}=
\sum\limits_{k=0}^N
\frac{(-i)^kH_k(\ell_m^{L,(p,q)} /\sqrt{2})}
{{\cal N} \sqrt{2^{k/2}[k]!}}
\CVector{N}{(p,q)}{k}.
\label{eq:harmonic}
\end{equation}
The above vector elements constitute  the analogue corresponding
to the  basis of ``sphe\-rical harmonic'' functions of the usual oscillator.

\subsection{ Summary}

In conclusion, the two-dimensional anisotropic quantum harmonic oscillator
with rational ratio of frequencies equal to $m/n$, is described dynamically
by a deformed version of the u(2) Lie algebra, the order of this algebra
being $m+n-1$. The representation modules of this algebra can be generated
by using the deformed oscillator algebra. The energy eigenvalues are
calculated by the requirement of the existence of finite dimensional
representation modules. An ``angular momentum'' operator useful for
labelling degenerate states has also been constructed.

The extension of the present method to the three-dimensional anisotropic
quantum harmonic oscillator is already receiving attention, since it is of
clear interest in the study of the symmetries underlying the structure of
superdeformed and hyperdeformed nuclei.



\section{References}


\def\ref{\par\hangindent=1.0cm\hangafter=1}
\parindent=0pt

\ref [1]  V. G. Drinfeld, in {\it Proceedings of
the International Congress of Mathematicians}, ed. A. M. Gleason
(American Mathematical Society, Providence, RI, 1986), p. 798.

\ref [2]  M. Jimbo, {\it Lett. Math. Phys.} {\bf 11} (1986) 247.

\ref [3]  L. C.  Biedenharn, {\it J. Phys. A} {\bf 22} (1989) L873.

\ref [4]  A. J. Macfarlane, {\it J. Phys. A} {\bf 22} (1989) 4581.

\ref [5]  C. P. Sun and H. C. Fu, {\it J. Phys. A} {\bf 22} (1989)
 L983.

\ref [6] M.  Arik and D. D. Coon, {\it J. Math. Phys.} {\bf 17} (1976)
 524.

\ref [7] V. V. Kuryshkin, {\it  Annales de la Fondation Louis de Broglie}
{\bf 5} (1980) 111.

\ref [8]  C. Daskaloyannis, {\it J. Phys. A} {\bf 24} (1991) L789.

\ref [9]  D.  Bonatsos and C. Daskaloyannis, {\it Phys. Lett. B}
 {\bf 307} (1993) 100.

\ref [10]  D. Bonatsos, C. Daskaloyannis, and P. Kolokotronis, {\it J.
Phys. A} {\bf 26} (1993) L871.

\ref [11]  C. Delbecq and C. Quesne, {\it J. Phys. A} {\bf 26}
 (1993) L127.

\ref [12]  F. Pan, {\it J. Math. Phys.} {\bf 35} (1994) 5065.

\ref [13]  P. P. Raychev, R. P. Roussev and Yu. F. Smirnov, {\it J.
Phys. G} {\bf 16} (1990) L137.

\ref [14] D. Bonatsos, E. N. Argyres, S. B. Drenska, P. P. Raychev,
R. P. Roussev and Yu. F. Smirnov, {\it Phys. Lett. B} {\bf 251} (1990) 477.

\ref [15] D. Bonatsos, S. B. Drenska, P. P. Raychev, R. P. Roussev and
Yu. F. Smirnov, {\it J. Phys. G} {\bf 17} (1991) L67.

\ref [16] D. Bonatsos, A. Faessler, P. P. Raychev, R. P. Roussev and
Yu. F. Smirnov, {\it J. Phys. A} {\bf 25} (1992) 3275.

\ref [17] D. Bonatsos, C. Daskaloyannis, A. Faessler, P. P. Raychev
and R. P. Roussev, {\it Phys. Rev. C} {\bf 50} (1994) 497.

\ref [18] P. Holmberg and P. Lipas, {\it Nucl. Phys. A} {\bf 117} (1968) 552.

\ref [19] D. Bonatsos, {\it J. Phys. A} {\bf 25} (1992) L101.

\ref [20] D. Bonatsos and C. Daskaloyannis, {\it Phys. Lett. B}
{\bf 278} (1992) 1.

\ref [21] D. Bonatsos, C. Daskaloyannis and A. Faessler, {\it J. Phys. A}
{\bf 27} (1994) 1299.

\ref [22] D. Bonatsos, A. Faessler, P. P. Raychev, R. P. Roussev
and Yu. F. Smirnov, {\it J. Phys. A} {\bf 25} (1992) L267.

\ref [23] D. P. Menezes, S. S. Avancini and C. Provid\^encia,
{\it J. Phys. A} {\bf 25} (1992) 6317.

\ref [24] D. Bonatsos, L. Brito, D. P. Menezes, C. Provid\^encia and
J. da Provid\^encia, {\it J. Phys. A} {\bf 26} (1993) 895, 5185.

\ref [25] C. Provid\^encia, L. Brito, J. da Provid\^encia, D. Bonatsos
and D. P. Menezes, {\it J. Phys. G} {\bf 20} (1994) 1209.

\ref [26] B. Mottelson, {\it Nucl. Phys. A} {\bf 522} 1c.

\ref [27] W. D. M. Rae, {\it Int. J. Mod. Phys. A} {\bf 3} (1988) 1343.

\ref [28] W. D. M. Rae and J. Zhang, {\it Mod. Phys. Lett. A} {\bf 9}
(1994) 599.

\ref [29] D. M. Brink, in {\it Proc. Int. School of Physics, Enrico
Fermi Course XXXVI, Varenna 1966}, ed. C. Bloch (Academic Press, New York,
1966),  p. 247.

\ref [30] T. P. Martin, T. Bergmann, H. G\"ohlich and T. Lange,
{\it Z. Phys. D} {\bf 19} (1991) 25.

\ref [31] A. Bulgac and C. Lewenkopf, {\it Phys. Rev. Lett. } {\bf 71}
(1993) 4130.

\ref [32] D. Bonatsos, P. P. Raychev, R. P. Roussev and Yu. F. Smirnov,
{\it Chem. Phys. Lett. } {\bf 175} (1990) 300.

\ref [33] D. Bonatsos, E. N. Argyres and P. P. Raychev, {\it J. Phys. A}
{\bf 24} (1991) L403.

\ref [34] D. Bonatsos, P. P. Raychev and A. Faessler, {\it Chem. Phys. Lett.}
{\bf 178} (1991) 221.

\ref [35] D. Bonatsos and C. Daskaloyannis, {\it Phys. Rev. A} {\bf 46}
(1992) 75.

\ref [36] D. Bonatsos, C. Daskaloyannis and K. Kokkotas, {\it J. Math. Phys.}
{\bf 33} (1992) 2958.

\ref [37] D. Bonatsos, C. Daskaloyannis and H. A. Mavromatis, {\it Phys.
Lett. A} {\bf 199} (1995) 1.

\ref [38] R. N. Alvarez, D. Bonatsos and Yu. F. Smirnov, {\it Phys.
Rev. A} {\bf 50} (1994) 1088.

\ref [39] D. Bonatsos and C. Daskaloyannis, {\it Chem. Phys. Lett.}
{\bf 203} (1993) 150.

\ref [40] D. Bonatsos and C. Daskaloyannis, {\it Phys. Rev. A} {\bf 48}
(1993) 3611.

\ref [41]  J. M. Jauch and E. L. Hill, {\it Phys. Rev.} {\bf 57}
 (1940) 641.

\ref [42]  G. Contopoulos, {\it Z. Astrophys.} {\bf 49} (1960) 273;
{\it Astrophys. J.} {\bf 138} (1963) 1297.

\ref [43]  G. Maiella and G. Vilasi, {\it Lettere Nuovo Cimento}
 {\bf 1} (1969) 57.

\ref [44]  W. Nazarewicz and J. Dobaczewski, {\it Phys. Rev. Lett.}
 {\bf 68} (1992) 154.

\ref [45]  J. Hietarinta, {\it Phys. Rep.} {\bf 147} (1987) 87.

\ref [46] C. R. Holt, {\it J. Math. Phys.} {\bf 23} (1982) 1037.

\ref [47] A. S. Fokas and P. A. Lagerstrom, {\it J. Math. Anal. Appl. }
{\bf 74} (1980) 325.

\ref [48] D.  Bonatsos, C.  Daskaloyannis and K. Kokkotas, {\it Phys. Rev.
A} {\bf 48}  (1993) R3407; {\bf 50} (1994) 3700.

\ref [49]  P. W. Higgs, {\it J. Phys. A} {\bf 12} (1979) 309.

\ref [50]  R. Floreanini, J. LeTourneux and L. Vinet, {\it Ann. Phys.}
{\bf 226} (1993) 331.

\ref [51]  D. Bonatsos, C.  Daskaloyannis, D. Ellinas and
A. Faessler, {\it Phys. Lett. B} {\bf 331} (1994) 150.

\ref [52] A. A. Kehagias and G. Zoupanos, {\it Z. Phys. C} {\bf 62}
(1994) 121.


\end{document}